\documentclass[reprint,amssymb,amsmath,aip,cha]{revtex4-2}
\usepackage{bm}%
\usepackage{color}
\usepackage{graphicx}
\usepackage{mathptmx}
\usepackage{amsmath}
\usepackage{derivative}
\usepackage[utf8]{inputenc}
\usepackage[T1]{fontenc}
\DeclareUnicodeCharacter{2009}{\,}

\begin{document}

\title{Magnonic Fabry-P\'{e}rot resonators as programmable phase shifters}

\author{Anton Lutsenko}
\affiliation{NanoSpin, Department of Applied Physics, Aalto University School of Science, PO Box 15100, FI-00076 Aalto, Finland}

\author{Kevin G. Fripp}
\affiliation{University of Exeter, Stocker Road, Exeter, EX4 4QL, United Kingdom}

\author{Luk\'{a}\v{s} Flaj\v{s}man}
\affiliation{NanoSpin, Department of Applied Physics, Aalto University School of Science, PO Box 15100, FI-00076 Aalto, Finland}

\author{Andrey V. Shytov}
\author{Volodymyr V. Kruglyak}\email[Corresponding author: ]
{v.v.kruglyak@exeter.ac.uk}
\affiliation{University of Exeter, Stocker Road, Exeter, EX4 4QL, United Kingdom}

\author{Sebastiaan van Dijken}
\email[Corresponding author: ]
{sebastiaan.van.dijken@aalto.fi}
\affiliation{NanoSpin, Department of Applied Physics, Aalto University School of Science, PO Box 15100, FI-00076 Aalto, Finland}


\begin{abstract}
We explore the use of magnonic Fabry-Pérot resonators as programmable phase shifters for spin-wave computing. The resonator, composed of a yttrium iron garnet (YIG) film coupled with a CoFeB nanostripe, operates through dynamic dipolar coupling, leading to wavelength downconversion and the formation of a magnonic cavity. Using super-Nyquist sampling magneto-optical Kerr effect (SNS-MOKE) microscopy and micromagnetic simulations, we demonstrate that these resonators can induce a $\pi$ phase shift in the transmitted spin wave. The phase shift is highly sensitive to the magnetization alignment within the resonator, allowing for on-demand control via magnetic switching. This feature, combined with low-loss transmission, positions the magnonic Fabry-Pérot resonator as a promising component for reconfigurable magnonic circuits and spin-wave computing devices.
\end{abstract}

\maketitle

Magnonics holds substantial promise for advancing data processing through improvements in energy efficiency, processing speed, and operational functionality\cite{Barman2021,Chumak2022}. In magnonics, information is encoded in the amplitude or phase of spin waves propagating through magnetic films\cite{Khitun2010,Mahmoud2020}. A notable example of a phase-encoded device is the spin-wave majority gate\cite{Khitun2011,Fischer2017,Talmelli2020}. The operation of this logic gate relies on the interference of three spin waves of equal amplitude, with information encoded in their phases. These phases can be equal to or differ by $\pi$, causing the phase of the output wave to reflect the majority phase of the inputs. A programmable phase shifter that can modify the phase of propagating spin waves is an essential component for efficient phase-based information processing. 

Researchers have demonstrated various tunable spin-wave phase shifters based on different operational principles. Among these, nanoscale chiral magnonic resonators, which interact with propagating spin waves via dynamic dipolar fields, stand out\cite{Au2012,Fripp2021}. These devices produce frequency-dependent adjustments to the spin-wave amplitude and phase that depend on the spin-wave propagation direction and the alignment of magnetization within the resonator\cite{Au2012,Fripp2021}. Other methods for phase modulation include interactions with magnetic domain walls\cite{Hertel2004}, Oersted fields generated by current-carrying wires\cite{Demidov2009}, magnetic defects\cite{Baumgaertl2018}, spin currents\cite{Zhang2020}, regions of inverted magnetization\cite{Baumgaertl2021}, and voltage-controlled magnetic domains\cite{Qin2021}. Nonlinear effects at high excitation powers can also induce phase shifts\cite{Ustinov2021,Lake2022}.

For successful integration of spin-wave phase shifters into magnonic circuits, several criteria must be met. The phase-shifting elements must be scalable, capable of achieving significant phase modulation over short distances, and operate under low-power conditions to maximize energy efficiency. Additionally, in applications such as field programmable gate arrays and other reconfigurable devices, the phase shift should be controllable. The ability to switch the phase by $\pi$ on demand is a particularly desirable feature in wave-based computing architectures\cite{Khitun2010,Mahmoud2020}. 

The recently introduced magnonic Fabry-P\'{e}rot resonator holds promise for meeting these requirements\cite{Qin2021b,Talapatra2023}. This resonator consists of a low-loss magnetic medium, typically an yttrium iron garnet (YIG) film, coupled with a magnetic nanostripe. It operates in a frequency range where spin waves propagate in YIG but remain below the ferromagnetic resonance frequency (FMR) of the nanostripe. Through dynamic dipolar coupling, the resonator achieves nonreciprocal wavelength downconversion, forming a magnonic cavity where spin waves circulate between the edges of the bilayer. This configuration creates narrow gaps in the spin-wave transmission spectrum while maintaining minimal signal loss at intermediate frequencies, enabling low-loss spin-wave manipulation. The significant wavelength reduction within the resonator allows for nanoscale control of micrometer-scale waves.

\begin{figure*}
	\centering
	\includegraphics[width=2.0\columnwidth]{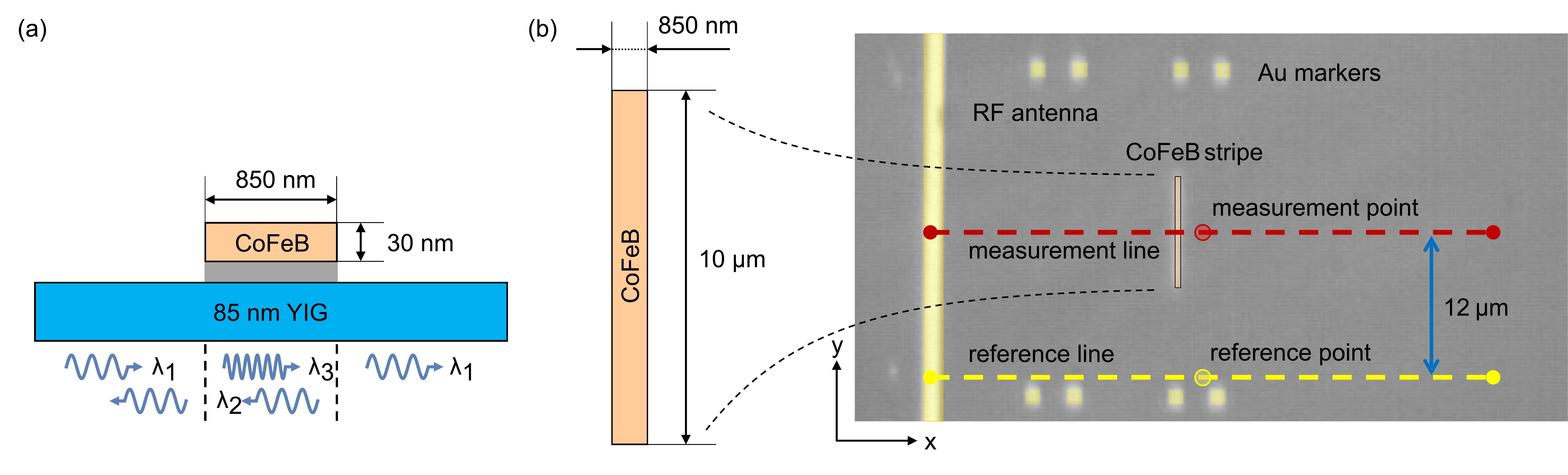}
	\caption{(a) Schematic of the magnonic Fabry-P\'{e}rot resonator structure. The resonator comprises a 30-nm-thick CoFeB nanostripe patterned on an 85-nm-thick YIG film. The nanostripe is 850 nm wide and 10 $\mu$m long. Dynamic dipolar coupling between the CoFeB nanostripe and the YIG film reduces the spin-wave wavelength in the bilayer region. Additionally, due to the chiral nature of dipolar coupling, spin waves propagating in opposite directions exhibit different wavelengths. The propagation directions of the waves with longer $\lambda_2$ and shorter $\lambda_3$ wavelengths depend on the magnetization orientation in the YIG film. The schematic illustrates the wavelength conversion occurring when the YIG film's magnetization is aligned along the positive $y$-direction. (b) Microscopy image of the sample along with the measurement geometry. The distance between the Fabry-P\'{e}rot resonator and the 1.5-$\mu$m-wide microwave antenna is 20 $\mu$m. We conducted SNS-MOKE microscopy along a measurement line passing through the center of the resonator (indicated by the red dashed line) and along a parallel reference line without the resonator (indicated by the yellow dashed line).}
	\label{Fig1}
\end{figure*}

Previous research on magnonic Fabry-P\'{e}rot resonators has demonstrated their effectiveness in amplitude modulation and wavelength conversion\cite{Qin2021b,Talapatra2023}. In this work, we investigate their potential as programmable phase shifters. Using super-Nyquist sampling magneto-optical Kerr effect (SNS-MOKE) microscopy and micromagnetic simulations, we show that these nanoscale resonators can induce a $\pi$ phase shift in the transmitted spin wave near the transmission gap frequency. Additionally, we demonstrate that the phase can be switched on demand by altering the magnetization alignment within the resonator between parallel and antiparallel configurations via an applied magnetic field.

Figure \ref{Fig1} shows the sample and measurement geometry. The sample consists of an 85-nm-thick YIG film with a CoFeB nanostripe patterned on top. We grew the YIG film, with a Gilbert damping parameter $\alpha=5\times10^{-4}$, on a (111)-oriented single-crystal Gd$_3$Ga$_5$O$_{12}$ (GGG) substrate using pulsed laser deposition (PLD) at room temperature. The film was crystallized through post-deposition annealing at 750$^\circ$C for 8 hours in 13 mbar oxygen. The CoFeB nanostripe, composed of a 40:40:20 alloy, was fabricated using electron-beam lithography, magnetron sputtering, and lift-off processes. It is 30 nm thick, 10 $\mu$m long, and 850 nm wide. The magnetizations of the CoFeB nanostripe and the YIG film couple through dynamic dipolar fields across a 1 nm Ta/3 nm TaO$_x$ spacer. 

Spin waves were excited in the YIG film using a gold microwave antenna positioned 20 $\mu$m from the Fabry-P\'{e}rot resonator. The antenna, measuring 120 nm in thickness, 1.5 $\mu$m in width, and 50 $\mu$m in length, was operated at an excitation power of 0 dBm. We applied a magnetic bias field parallel to the antenna to ensure the excitation of Damon-Eshbach spin waves in all experiments. Transmission characteristics of these spin waves through the Fabry-P\'{e}rot resonator were analyzed using SNS-MOKE microscopy in a home-built setup\cite{Dreyer2021,Qin2021}. To investigate the resonator's impact on the amplitude and phase of transmitted spin waves, we conducted SNS-MOKE scans along a horizontal line crossing the center of the CoFeB nanostripe and along another parallel line away from the resonator, as illustrated in Fig. \ref{Fig1}(b). 

We complemented our experiments with micromagnetic simulations using MuMax3 software\cite{Vansteenkiste2014}. For the 85-nm-thick YIG film, we used input parameters including a saturation magnetization $M_{s}=1.2\times10^5$ A/m, exchange constant $A_{ex}=3.5\times10^{-12}$ J/m, and a Gilbert damping parameter $\alpha=5\times10^{-4}$. The 30-nm-thick, 850-nm-wide CoFeB nanostripe was modeled with $M_{s}=1.15\times10^6$ A/m, $A_{ex}=1.6\times10^{-11}$ J/m, and $\alpha=5\times10^{-3}$. A 5-nm-thick spacer separated the CoFeB nanostripe from the YIG film. The simulation area was discretized using cuboidal cells with 5 nm edge lengths and a grid size of $16384\times2\times24$ in the $x$, $y$, and $z$ directions. The 10-$\mu$m-long CoFeB nanostripe was approximated as infinitely long with one-dimensional periodic boundary conditions along its length (the $y$-axis), which provides accurate results just behind the resonator. We continuously excited spin waves using a sinusoidal out-of-plane field at frequencies ranging from 1.0 GHz to 1.6 GHz. The spatial profile of the excitation field was simulated using COMSOL's Coils geometry to replicate the antenna used in the experiments at the same power level and position. This profile, defined in COMSOL on an irregular mesh, was interpolated onto the regular grid used in MuMax3 software. Simulations ran for 60 ns before data collection to ensure steady-state conditions. We sampled the time evolution of the film's magnetization behind the CoFeB nanostripe at 50 ps intervals, and increased the Gilbert damping parameter parabolically near the YIG film edges to prevent spin-wave back-reflection. The influence of the Fabry-P\'{e}rot resonator on the amplitude and phase of transmitted spin waves was determined by comparing these results with a reference simulation on a YIG film without the CoFeB nanostripe.  

\begin{figure*}
	\centering
	\includegraphics[width=2.0\columnwidth]{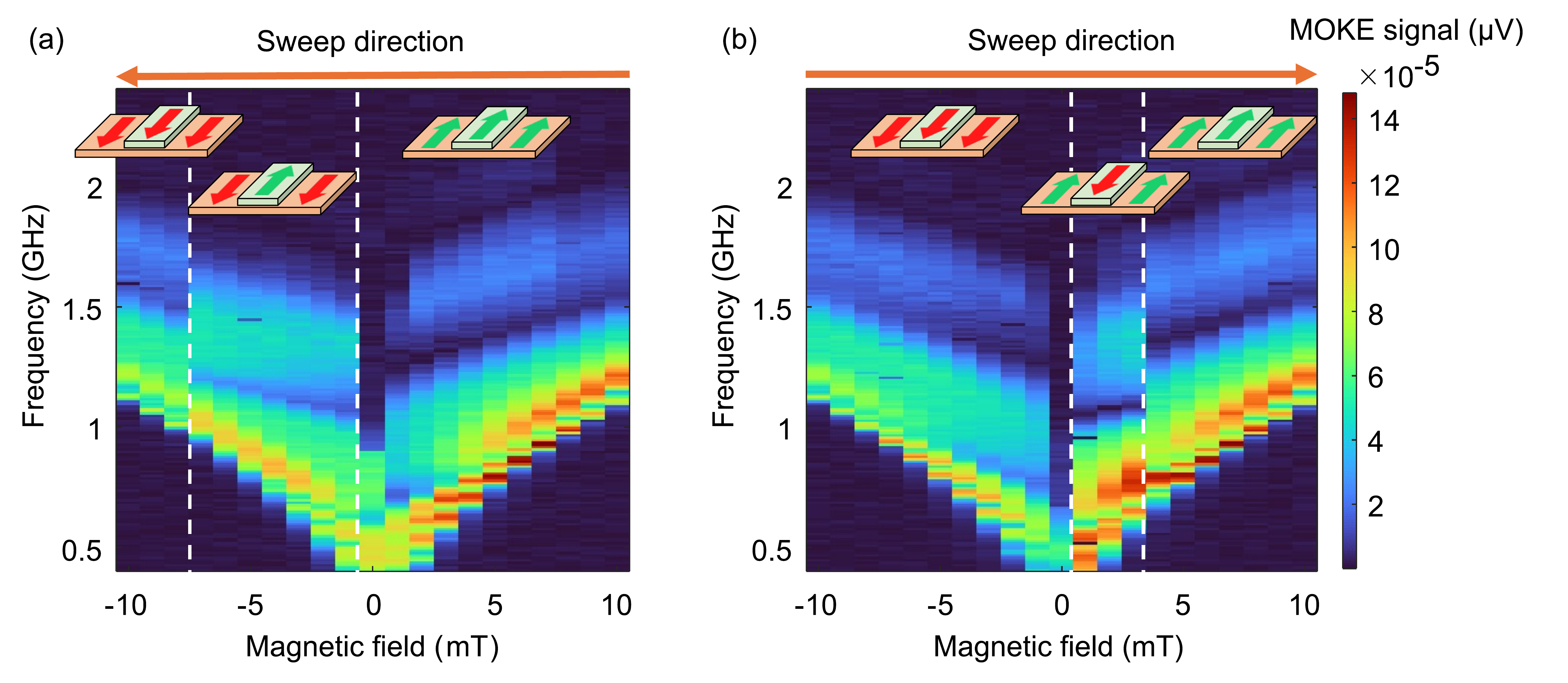}
	\caption{(a,b) SNS-MOKE signal as a function of the applied magnetic field and excitation frequency for two field-sweep directions, indicated by the orange arrows. Data are recorded 2 $\mu$m behind the center of the CoFeB nanostripe. The signal scales with the transmitted spin-wave amplitude. Vertical dashed lines mark the field values where abrupt changes in the frequency of the transmission gap occur. These changes correspond to switching between between parallel and antiparallel magnetization states within the resonator. The schematics illustrate the magnetization configuration in each field range.}
	\label{Fig2}
\end{figure*}

Figure \ref{Fig2} presents the SNS-MOKE signal measured 2 $\mu$m behind the center of the CoFeB nanostripe as a function of the applied magnetic field and excitation frequency. The signal scales with the oscillation amplitude of the $z$-component of magnetization ($m_z$), corresponding to the transmitted spin-wave amplitude. Distinct gaps are consistently observed in the transmission spectrum across the field range, with abrupt shifts in the gap frequency at certain field values. We attribute these gaps to the $n=2$ Fabry-P\'{e}rot resonance and their shifts to changes in the spin-wave dispersion within the resonator when its magnetization configuration switches between parallel and antiparallel states\cite{Qin2021b}. Specifically, as the field is reversed, the magnetizations of the YIG film and CoFeB nanostripe also reverse but at different field values, resulting in parallel and antiparallel relative orientations (shown in Fig. \ref{Fig2} insets) across different field ranges. Moreover, each parallel and antiparallel configuration yields two different transmission spectra, leading to four 'magnonic' states. This is a manifestation of the system's chirality\cite{Kruglyak2021}: the transmission depends on the spin wave's incidence direction, and so, the two magnetic states obtained by rotating the system (in any configuration) in the film plane by 180{\textdegree} (and so switching the direction of spin-wave propagation) must be distinguished. The gaps corresponding to the antiparallel states occur at lower frequencies than those for the parallel states. Our SNS-MOKE measurements on the 10-$\mu$m-long resonator align closely with previous broadband spin-wave spectroscopy results from studies on longer magnonic Fabry-P\'{e}rot resonators\cite{Qin2021b}, indicating that the spin-wave characteristics near the resonator are minimally impacted by its reduced length. 

\begin{figure}
	\centering
	\includegraphics[width=1.0\columnwidth]{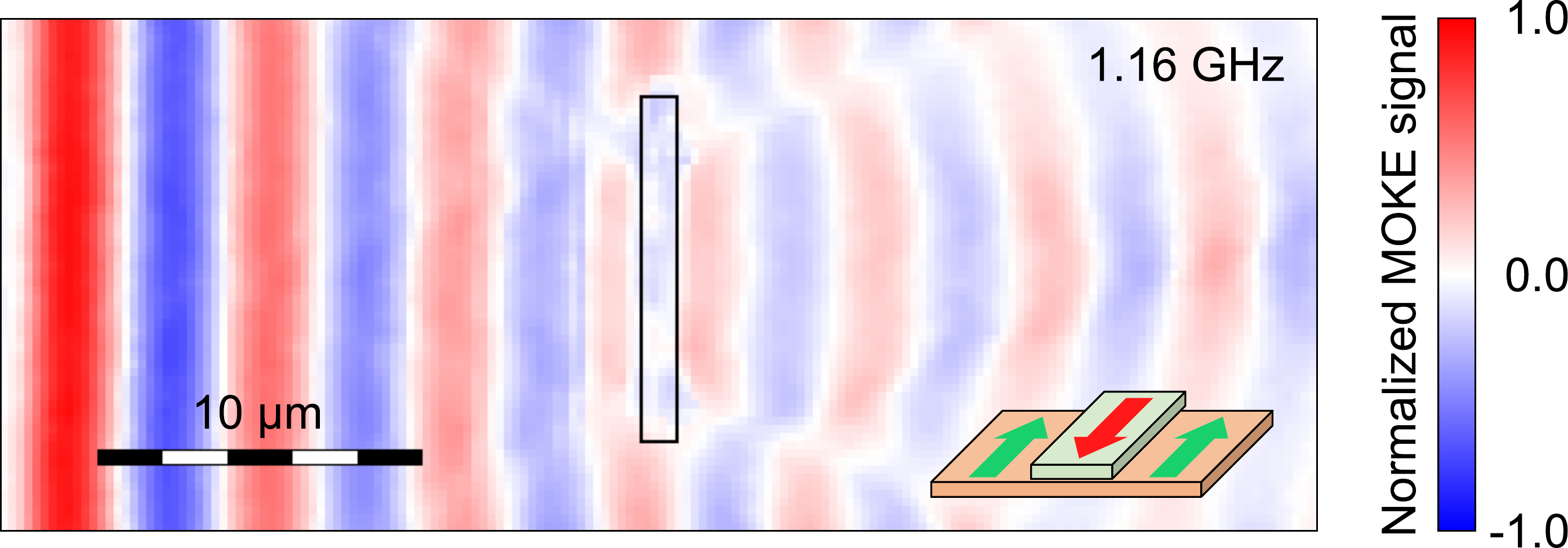}
	\caption{SNS-MOKE microscopy images depicting spin-wave transport across the 850-nm-wide and 10-$\mu$m-long Fabry-P\'{e}rot resonator at 1.16 GHz under a 3 mT field. The schematic illustrates the antiparallel magnetization configuration. Spin-wave transmission across the resonator induces a phase shift.}
    \label{Fig3}
\end{figure}

\begin{figure*} 
	\centering
	\includegraphics[width=1.8\columnwidth]{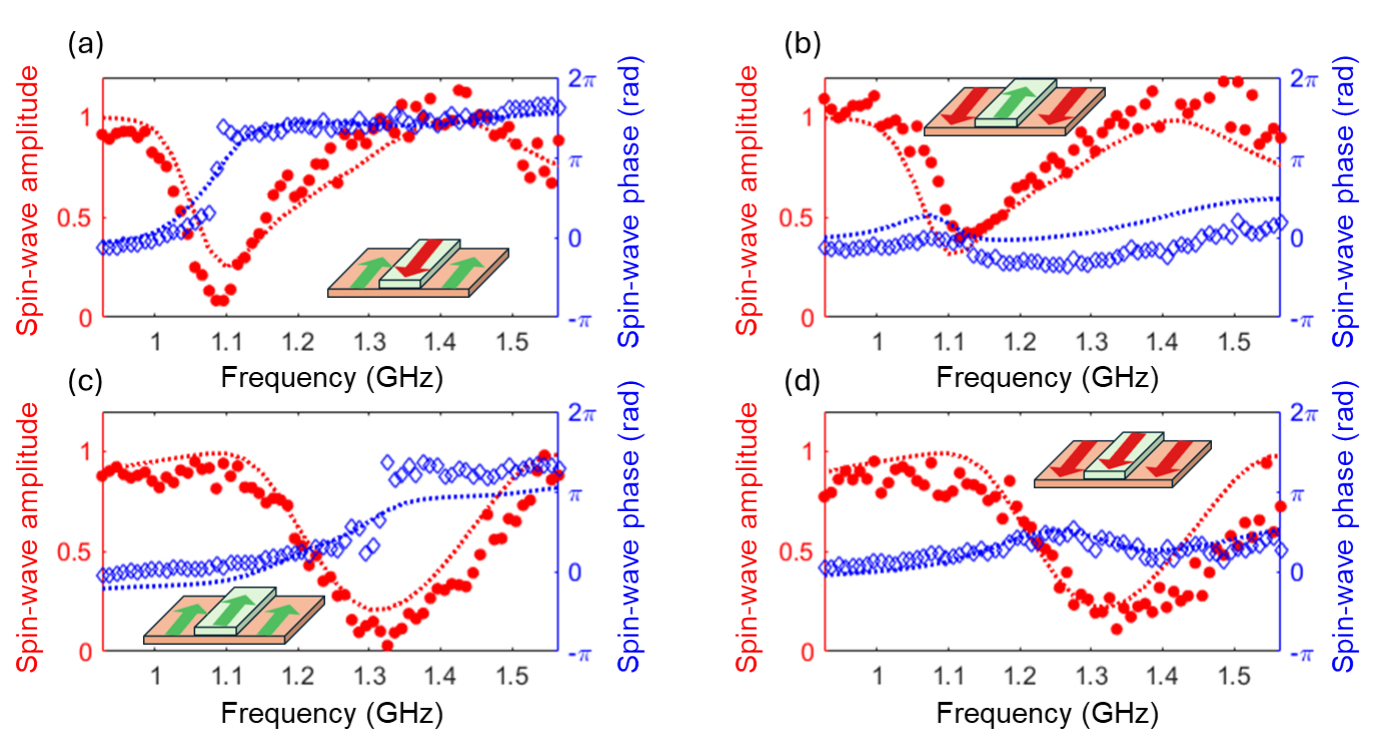}
   	\caption{Influence of the Fabry-P\'{e}rot resonator on the amplitude and phase of transmitted spin waves for the two distinct antiparallel (a,b) and parallel (c,d) magnetization configurations at a 3 mT field. The symbols indicate data from the SNS-MOKE measurements, while the lines represent results from the simulations. The spin waves are incident from the left, and the amplitude and phase information is derived 4.8 $\mu$m behind the resonator. The schematics illustrate the magnetization configuration.}
	\label{Fig4}
\end{figure*}

Figure \ref{Fig3} presents an SNS-MOKE microscopy image of the $z$-component of magnetization ($m_z$) for antiparallel magnetization at 1.16 GHz under a 3 mT field. This image shows how the parallel wavefront of the incident spin wave deforms when it interacts with the finite-sized resonator: the wavefront curves, and spin-wave caustic beams \cite{Veerakumar2006, Gieniusz2013, Davies2015} emanate from the nanostripe's edges in four directions. The curvature results from a significant phase shift produced by the resonator, accumulated over a short distance of 850 nm, i.e., the resonator width. This phase shift bends the wavefront to maintain phase continuity \cite{Whitehead2019}. As the spin wave propagates beyond the Fabry-P\'{e}rot resonator, both the phase shift and wavefront curvature gradually diminish.      

To explore the resonator's phase-shifting characteristics in detail, we compared the complex-valued SNS-MOKE signal recorded along a horizontal line across the center of the CoFeB nanostripe with that of a reference acquired along a parallel line away from the resonator (see Fig. \ref{Fig1}(b)). The spin-wave transmission coefficient was calculated as the ratio of these two signals, acquired 4.8 $\mu$m behind the resonator to avoid distortions from caustic beams (see Fig. \ref{Fig3}). Figure \ref{Fig4} shows the resulting amplitude and phase of the spin-wave transmission coefficient for the four magnetization configurations of the resonator under a 3 mT field, alongside micromagnetic simulation results for comparison.

\begin{figure*}
	\centering
	  \includegraphics[width=2\columnwidth]{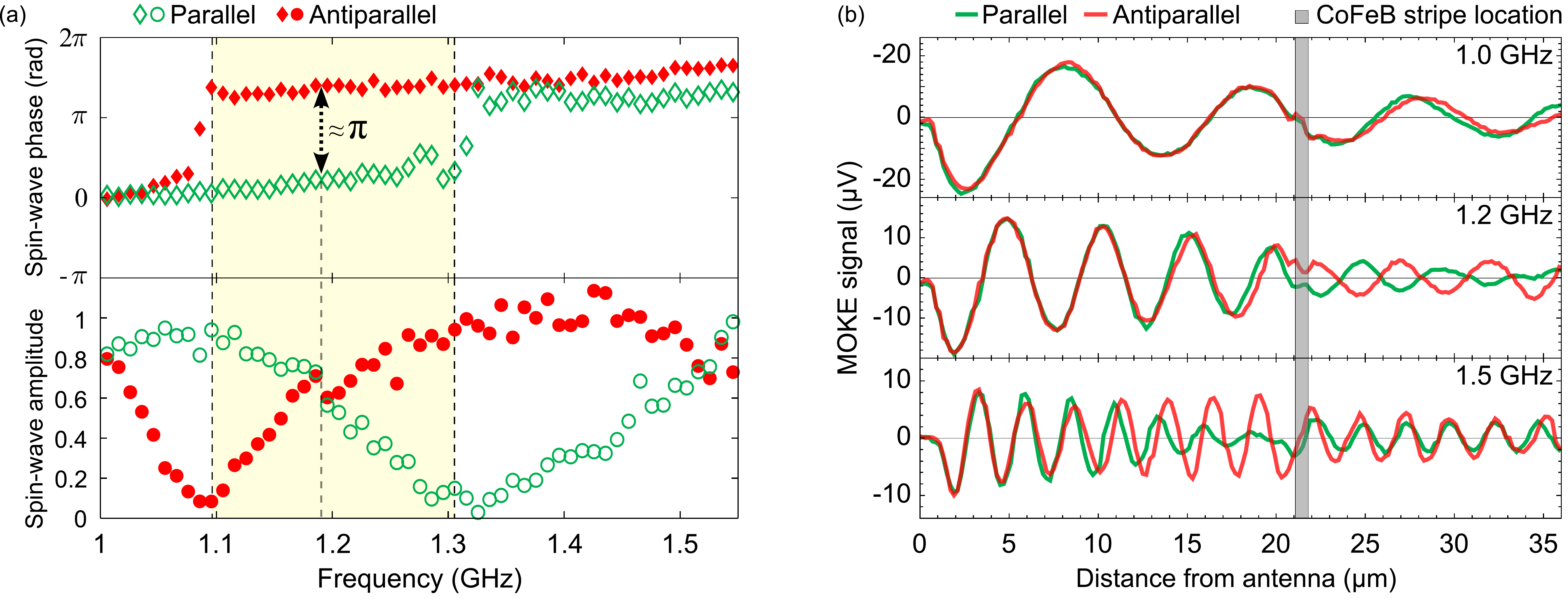}
	\caption{(a) Illustration of how magnetic switching of the magnonic Fabry-P\'{e}rot resonator induces a phase shift of $\pi$ while maintaining the amplitude of transmitted spin waves. Around 1.2 GHz, the spin-wave amplitude behind the resonator is identical for the antiparallel (red symbols) and parallel (green symbols) magnetization configuration (bottom panel), but the phase difference is $\pi$ (top panel). The experimental data are taken from Fig. \ref{Fig4}(a,c). (b) SNS-MOKE linescans recorded along a horizontal line across the center of the CoFeB nanostripe for the parallel and antiparallel magnetization configurations at 1.0 GHz, 1.2 GHz, and 1.5 GHz. The position of the CoFeB nanostripe is marked by a grey stripe. The two magnetization states transmit the spin wave similarly at 1.0 GHz and 1.5 GHz, but at 1.2 GHz, the phase of the transmitted signal shifts by $\pi$ upon magnetic switching in the resonator.}
	\label{Fig5}
\end{figure*}

When the YIG film magnetization aligns along the positive $y$-direction, the resonator significantly suppresses the the transmitted spin-wave amplitude at its Fabry-P\'{e}rot resonance frequency. The resonance frequency differs for the parallel and antiparallel configurations (Fig. \ref{Fig4}(a,c)). In both states, the incoming spin wave with wavelength $\lambda_1$ converts to a much shorter $\lambda_3$ wave at the first CoFeB/YIG bilayer edge, as shown in Fig. \ref{Fig1}(a). However, due to differences in the dispersion relations, the resonance frequency corresponding to the resonant wavelength varies substantially between the parallel and antiparallel magnetization states\cite{Qin2021b}. The transmitted spin-wave phase also varies with frequency. Near resonance, the phase shift induced by the Fabry-P\'{e}rot resonator evolves from about 0 to over $\pi$ as the frequency increases (Fig. \ref{Fig4}(a,c)). In the antiparallel configuration around 1.1 GHz, the phase shift changes more sharply with frequency than in the parallel configuration near 1.3 GHz, consistent with the narrower transmission gap in the antiparallel state. 

In contrast, when the YIG film magnetization aligns along the negative $y$-direction, the resonant modulation of the spin-wave amplitude and phase is weaker (Fig. \ref{Fig4}(b),(d)). In this configuration, the incoming spin wave with wavelength $\lambda_1$ first converts to a $\lambda_2$ wave at the first CoFeB/YIG bilayer edge and then reconverts to $\lambda_1$ at the second edge. We attribute the weaker signal modulation to the lower damping of the $\lambda_2$ wave within the resonator compared to the $\lambda_3$ wave, as shown in Refs. \citenum{Qin2021b} and \citenum{Talapatra2023}, as well as notable differences in the interface transmission coefficients for the $\lambda_1\rightarrow\lambda_2\rightarrow\lambda_1$ versus the $\lambda_1\rightarrow\lambda_3\rightarrow\lambda_1$ conversion processes. This interpretation is consistent with that from Ref. \citenum{Smigaj2023}. 

The Fabry-Pérot resonator's ability to induce distinct amplitude and phase modulation in parallel and antiparallel configurations enables reversible phase shifting through magnetic switching, particularly at intermediate frequencies. As illustrated in Fig. \ref{Fig5}(a), reversing the CoFeB nanostripe magnetization while maintaining the YIG film magnetization along the positive $y$-direction produces a phase shifts of approximately $\pm\pi$ in the 1.1 GHz to 1.3 GHz frequency range. In phase-based magnonic computing, achieving phase inversion while preserving the spin-wave amplitude is ideal. We accomplish this feature at a specific frequency, approximately 1.2 GHz, where the resonator decreases the spin-wave amplitude to about 65\% for both the parallel and antiparallel states, compared to the amplitude in the absence of the resonator. In other words, at this frequency and under 3 mT applied field, magnetic switching in the Fabry-P\'{e}rot resonator produces a $\pi$ phase shift with a minimal effect on the amplitude of the transmitted spin wave.

The SNS-MOKE linescans in Fig. \ref{Fig5}(b) further confirm the Fabry-P\'{e}rot resonator's functionality as a programmable phase shifter. At 1.0 GHz and 1.5 GHz, the resonator modulates the transmitted spin wave similarly in both magnetization states, producing nearly identical wave profiles beyond the CoFeB nanostripe. However, at 1.2 GHz, while the amplitude of the transmitted spin wave remains consistent, its phase shifts by $\pi$ upon magnetization switching in the resonator.   

In summary, this study demonstrates the effectiveness of a magnonic Fabry-P\'{e}rot resonator, composed of a YIG film coupled with a CoFeB nanostripe, as a programmable phase shifter. Our findings show that this resonator can achieve a $\pi$ phase shift while maintaining the transmitted spin-wave amplitude through magnetic switching. The resonator’s narrow 850 nm width provides a scalable solution to phase modulation, which is crucial for integrating such components into magnonic computing architectures where programmable phase shifting is essential for advanced data processing.\newline

This project has received funding from the European Union's Horizon Europe research and innovation program under Grant Agreement No. 101070347-MANNGA. Yet, views and opinions expressed are those of the authors only and do not necessarily reflect those of the EU, and the EU cannot be held responsible for them. The project also received funding from the UK Research and Innovation (UKRI) under the UK government’s Horizon Europe funding guarantee (Grant No. 10039217) as part of the Horizon Europe (HORIZON-CL4-2021-DIGITAL-EMERGING-01) under Grant Agreement No. 101070347.\\ 

\noindent \textbf{AUTHOR DECLARATIONS}

\noindent \textbf{Conflict of Interest}

The authors have no conflicts to disclose.\\
	
\noindent \textbf{Author Contributions}

A.L. performed the experiments. K.G.F. conducted the micromagnetic simulations. L.F. assisted in the SNS-MOKE measurements. V.V.K and S.v.D supervised the work. All authors discussed the results and contributed to the writing of the manuscript.\\ 
	
\noindent \textbf{DATA AVAILABILITY}

The data that support the findings of this study are available from the corresponding authors upon reasonable request.\\

\noindent \textbf{REFERENCES}
\providecommand{\noopsort}[1]{}\providecommand{\singleletter}[1]{#1}%

\end{document}